# Massive Parallelization of STED Nanoscopy Using Optical Lattices

Bin Yang[1,2], Frédéric Przybilla[1,2], Michael Mestre[1,2], Jean-Baptiste Trebbia[1,2] and

Brahim Lounis[1,2]*

[1]*Univ Bordeaux, LP2N, F-33405 Talence, France.*

[2]*Institut d'Optique & CNRS, LP2N, F-33405 Talence, France.*

* blounis@u-bordeaux1.fr

**Recent developments in stimulated emission depletion (STED) microscopy achieved nanometer scale resolution[1,2] and showed great potential in live cell imaging[3,4]. Yet, STED nanoscopy techniques[5] are based on single point-scanning. This constitutes a drawback for wide field imaging, since the gain in spatial resolution requires dense pixelation and hence long recording times. Here we achieve massive parallelization of STED nanoscopy using wide-field excitation together with well-designed optical lattices for depletion and a fast camera for detection. Acquisition of large field of view super-resolved images requires scanning over a single unit cell of the optical lattice which can be as small as 290 nm * 290 nm. Interference STED (In-STED) images of 2.9 µm * 2.9 µm with resolution down to 70 nm are obtained at 12.5 frames per second. The development of this technique opens many prospects for fast wide-field nanoscopy.**

Super-resolution microscopies based on single molecule localization[6-12] and saturated structured illumination[13-16] are intrinsically parallelized because they use wide-field illumination and cameras for detection. They have, however, very limited imaging speed. The first techniques require a low density of simultaneously emitting single molecules and therefore need a large number of frames for super-resolved image reconstruction. The second rely on sophisticated data post-processing, and consequently require high signal to noise camera frames accumulated over long integration times.

Freed from the constraint of sequentiality of single molecules emission, STED nanoscopy offers better temporal resolution[17]. However, STED and more generally RESOLFT[18,19] (REversible Saturable OpticaL Fluorescence Transitions) remain point-scanning techniques, which need parallelization in order to fully benefit from this temporal resolution for fast wide field imaging.

A first approach to STED parallelization is based on focused beam multiplication. A configuration using 4 pairs of scanning excitation and doughnut-STED beams, together with four avalanche photodiodes has been recently reported[20]. However, scaling the parallelization further up with this approach is mainly limited by the power of the available laser sources. Here, we show that well-designed optical lattices[21-25] can provide efficient depletion patterns for massive STED nanoscopy parallelization with moderate laser power.

Our In-STED microscopy setup, sketched in figure 1, is based on two synchronized laser sources delivering excitation and depletion pulses, akin to a standard STED microscope[26]. The laser beams are, however, not tightly focused but illuminate a wide-field region of the sample. A spatial Light Modulator (SLM) or a set of two Wollaston prisms is used to generate multiple depletion beams (3 or 4 respectively) which are sent on a high numerical aperture objective. An optical lattice is produced by the interference of the depletion beams at the sample plane and is overlaid with a uniform excitation beam. The fluorescence filtered from excitation and depletion photons is recorded using a fast CMOS camera.

In the 3 beam configuration (figure 2), the beams are parallel to the optical axis of the objective, and intersect its back focal plane at the vertices of a centered equilateral triangle. The optical lattice depends on the beams' polarization, and on $\theta$, the angle formed by the optical axis and the beams emerging from the objective. If the beams have the same polarization, parallel to one of the three sides of the triangle, their interference produces an hexagonal lattice with a periodicity $2\lambda/(3n \sin\theta)$, where $\lambda$ is the depletion laser wavelength and $n\sim1.5$ the sample refractive index. The intensity profile of the optical lattice, calculated for $\theta = 60°$ and displayed in figure 2b, shows a suitable depletion pattern for STED parallelization: an array of zero intensity minima, each of which being surrounded by a nearly uniform high intensity region.

To record the fluorescence depletion pattern, we scan a big fluorescent bead (>100 nm) with a piezo-stage over the field of illumination, while recording its fluorescence with the CMOS camera. For each scanning step, an image is acquired and the fluorescence intensity, integrated over the point spread function (PSF) of the bead image, is plotted in figure 2b. As expected, we obtain a hexagonal lattice with periodicity of 390 nm. The signal maxima of figure 2b correspond to the regions of minimal depletion, which occur at the zero-intensity positions of the optical lattice.

A key point for faster In-STED image acquisition is to generate lattices with the smallest possible unit cell. For this purpose, one can increase the angle $\theta$, but the aperture of the back focal plane of the objective sets a limit (figure 2). Alternatively, one can use a 4 beam configuration where a smaller unit cell can be obtained with the same angle $\theta$. In figure 2, two pairs of beams (one pair of beams polarized along the x axis and the other pair along the y axis) interfere independently and create a square optical lattice (figure 2c) with a periodicity $\lambda/(2n \sin\theta)$. Instead of an SLM, the beams can be generated using a combination of two Wollaston prisms in order to avoid the power losses inherent with non-ideal diffraction. Figure 2d shows the fluorescence depletion pattern obtained by scanning a large fluorescent bead for $\theta\sim60°$. This pattern has a periodicity of 290 nm, smaller than the one obtained in the previous configuration. Better parallelization also requires the largest depletion pattern possible. The size of this pattern is set as a compromise between the desired In-STED resolution and the available depletion power. Figure 2d shows that a 2.9 µm x 2.9 µm field of view can be obtained with efficient depletion using an average laser power of 400 mW. It corresponds to 100 unit cells of optical lattice and therefore a two order of magnitude gain in scanning area.

In-STED images of various samples are obtained as follows (Figure 3a): First we scan the sample over a unit cell in the presence of the wide field excitation and the depletion pattern, while acquiring a fluorescence image (128 * 128 pixels) for each scanning step. We then overlay a binary mask on these images, the transparent parts corresponding to the minima positions of the depletion pattern. Therefore, the CMOS camera together with the digital mask act as an array of parallelized "point detectors" (100 detectors, each giving the integrated signal of 13 camera pixels and recording an image of the size of the lattice unit cell). The complete In-STED image is then obtained by assembling all the unit cell images together.

Figures 3b and 3c display 2.9 µm * 2.9 µm images of a sample containing 20 nm fluorescent beads spin coated on a glass coverslip without and with the 4 beam depletion pattern (images obtained with the 3 beam configuration are presented in the supplementary information). Figure 3b clearly shows that the PSF of the beads are diffraction limited ~290 nm, while the resolution of the In-STED image of Figure 3c is well below this limit (typically ~70 nm). Since this resolution depends on the depletion intensity, it is better in the center of the pattern where the intensity is maximal. The resolution is ~1.5 less at the edges of the images where the intensity maxima are two times lower (the resolution scales as the inverse square root of the STED beam intensity [2]).

We apply In-STED microscopy to image microtubules in fixed COS cells. The microtubules were stained using a standard immunofluorescence protocol involving a primary antibody (anti beta-tubulin) and a secondary antibody labeled with fluorescent Dyes (Atto647N). As showed in the figure3 e-g, the In-STED gives super-resolved images of tubulin fibers. The resolution is below 100 nm and microtubules distant by less than the diffraction limit are clearly distinguished.

Interestingly, the acquisition time of an In-STED image ~80 ms is comparable to that obtained with fast-STED[17]. Unlike the latter, which is at the detected photons limit, In-STED is only limited by the CMOS camera readout time (summed over the frame stack of the unit cell scan) and can therefore be shortened further using faster cameras.

The In-STED setup provides many advantages compared to the multi-doughnut configuration[20]. First, with few interfering beams and a single chip detector, we can easily obtain a large number of intensity minima and their corresponding "point detectors". This is realized with a simple optical setup, with straightforward alignment of the excitation beam and optical lattice.

Secondly, for the same number of local minima and the same target resolution, the In-STED setup requires up to six-fold less depletion power than a multi-doughnut STED setup (see supplementary information). This is due to the fact that with interference one can achieve depletion intensities higher than that obtained for isolated donuts, and better-confined zero-intensity regions. Finally, the periodicity of the optical lattice is much smaller than the distance between the two neighbor doughnuts (several microns), therefore small scan regions and shorter acquisition times are required.

In conclusion, we showed that tailored optical lattices allow massive parallelization of standard STED microscopy. Super-resolved 2.9 µm x 2.9 µm images are obtained up to a rate of 12.5 frames per second, limited only by the CMOS camera readout time. A larger field of view can be achieved using a depletion laser with a lower repetition rate[27], which would provide higher intensity pulses for depletion. In-STED nanoscopy can easily be extended to other RESOLFT techniques based on photoactivable molecules[19], which require low intensities for photo-activation.

**Acknowledgements:** We warmly thank Philippe Tamarat for his valuable help with the laser sources, Olivier Rossier and Grégory Giannone for providing the cell culture and staining, and Laurent Cognet for helpful discussions. We acknowledge financial support from the Agence Nationale de la Recherche, Région Aquitaine, the French Ministry of Education and Research, the European Research Council and FranceBioImaging (Grant N° ANR-10-INSB-04-01).

**Methods:**

The fluorescence excitation beam at 571 nm is delivered by a frequency doubled optical parametric oscillator pumped by picosecond Ti-sapphire laser at a repetition rate of 76 MHz. A second Ti-sapphire laser, synchronized with the excitation laser and emitting at 760 nm, provides the depletion beams. The depletion laser average power is ~400 mW and the pulse duration is set to ~100 ps. A set of two dichroic mirrors is used to combine the excitation and depletion beams, and to separate the fluorescence. The fluorescence emission is collected with a high numerical aperture objective (60x NA=1.49), filtered from excitation and depletion photons and sent to a fast CMOS camera (ORCA-Flash4.0, HAMAMATSU). The SLM (or the set of two Wollaston prisms) is conjugated with the sample plane with a set of 3 lenses and the objective. We choose the total magnification of the microscope to be 225 so that the PSF spreads over more than 100 pixels (a pixel of the camera corresponds to ~29 nm at the sample plane). This increases the overall detection dynamics (higher detection saturation) and provides a better spatial discretization of the PSF to ensure minimal cross talk between the "point detectors". The number of pixels per "point detector" is chosen such that the ratio of the number of photons measured by the central detector to that measured by a neighboring one is less than 2%.

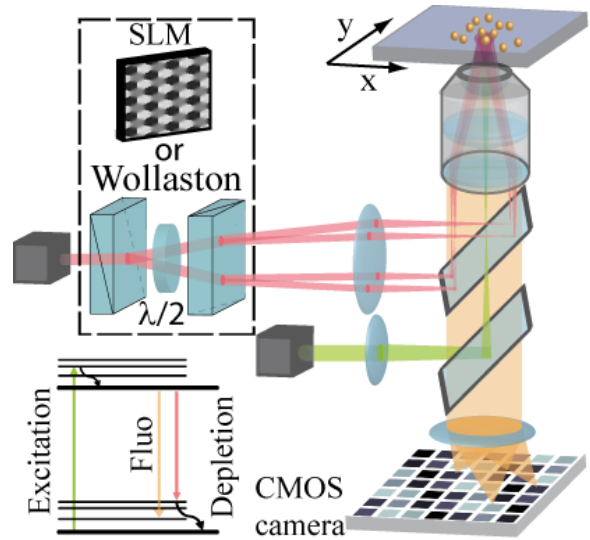

**Figure 1: In-STED experimental setup:** A depletion beam (red) is split with an SLM into 3 or 4 beams, or with a combination of two Wollaston prisms into 4 beams. The beams sent through an objective interfere at the sample plane to form an optical lattice. An excitation beam (green) of ~10 μm * 10 μm size is overlaid with the lattice. A CMOS camera conjugated with the sample plane records the wide-field fluorescence images.

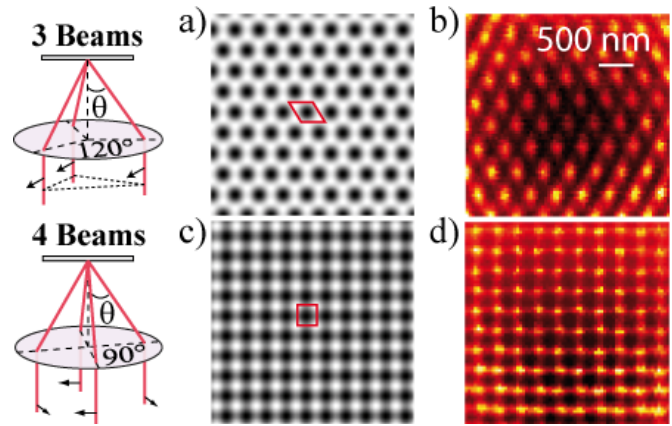

**Figure 2: Design of the optical lattices. Top left panel,** 3 beams, parallel to the optical axis of the objective, intersect its back focal plane at the vertices of a centered equilateral triangle. The beams have the same linear polarization, parallel to one of the three sides of the triangle. After passing through the objective they are deviated towards the focal region by an angle θ where they interfere. **a,** Intensity profile of the hexagonal optical lattice calculated for θ = 60° showing a suitable depletion pattern for STED parallelization. **b,** Fluorescence depletion pattern recorded with a 100 nm fluorescent bead scanned in the interference pattern (2.8 µm * 2.8 µm) in the presence of the excitation beam. **Bottom left panel,** The 4 beam configuration obtained with two pairs of beams (one pair of beams polarized along the x axis and the other pair along the y axis) interfere independently. **c,** Intensity profile of the obtained square optical lattice calculated for the same θ as in **a**. **d,** the corresponding depletion pattern.

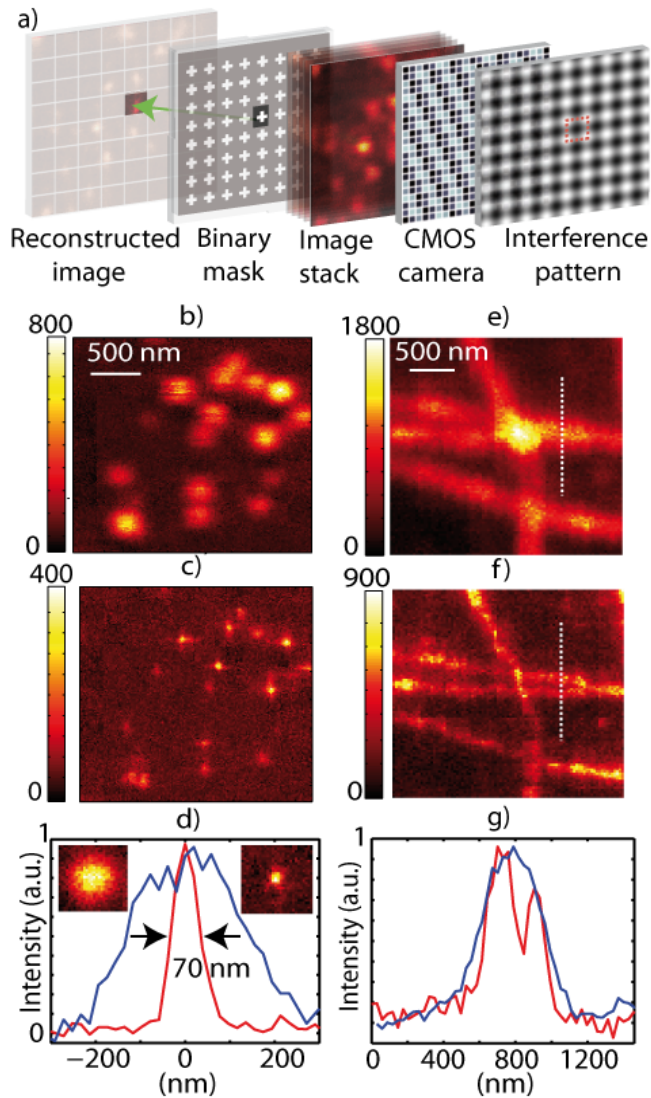

**Figure 3 : In-STED Images Acquisition: a,** a stack of a CMOS camera frames is acquired while the sample is scanned over a 290 nm x 290 nm optical lattice unit cell. A binary mask is applied to the frames. The "point-detectors" correspond to the white crosses *on* the mask, each of which records a unit cell image of the sample. **b,** diffraction limited image of 20 nm fluorescent beads recorded without the depletion beam at an excitation power of 2 mW (15 x 15 points per unit cell) **c,** superresolved In-STED image of the same region in the presence of the depletion beam (total depletion power ~400 mW). **d,** Normalized fluorescence intensity profiles measured from a single bead. **e,** diffraction limited image of microtubules in a fixed cell (10 x 10 points per scan area, integration time 800 μs per point). **f,** In-STED image of the same region. **g,** Normalized fluorescence intensity profiles (Cut along the dashed lines).